\documentclass[aps,pra,twocolumn,amsmath,amssymb,floatfix,superscriptaddress]{revtex4-2}
\usepackage{graphicx}
\usepackage[normalem]{ulem}
\usepackage[title]{appendix}
\usepackage{hyperref}
\usepackage{booktabs}
\usepackage{longtable}
\usepackage{multirow}
\usepackage{siunitx}

\makeatletter
\def\tableautorefname{Tab.\hskip\z@skip}
\def\subtableautorefname{Tab.\hskip\z@skip}
\def\figureautorefname{Fig.\hskip\z@skip}
\def\subfigureautorefname{Fig.\hskip\z@skip}
\def\sectionautorefname{Sec.\hskip\z@skip}
\def\subsectionautorefname{Sec.\hskip\z@skip}
\def\subsubsectionautorefname{Sec.\hskip\z@skip}
\def\appendixautorefname{App.\hskip\z@skip}
\def\chapterautorefname{Chapter\hskip\z@skip}
\def\equationautorefname~#1\null{%
  Eq.\hskip\z@skip~(#1)\null
}
\makeatother

\newcommand{\env}[1]{\texttt{#1}}
\usepackage[usenames,dvipsnames]{color}

\newcommand{\affA}{Van der Waals-Zeeman Institute, Institute of Physics,
University of Amsterdam, 1098 XH Amsterdam, Netherlands}
\newcommand{\affB}{QuSoft, Science Park 123, 1098 XG Amsterdam, the Netherlands}

\begin{document}

\title{Single ion spectroscopy of four metastable state clear-out transitions in Yb II: isotope shifts and hyperfine structure}

\author{N.~A.~Diepeveen}\affiliation{\affA}
\author{C.~Robalo Pereira}\affiliation{\affA} 
\author{M.~Mazzanti}\affiliation{\affA}
\author{Z.~E.~D.~Ackerman}\affiliation{\affA}
\author{L.~P.~H.~Gallagher}\affiliation{\affA}
\author{T.~Timmerman}\affiliation{\affA}
\author{R.~Gerritsma}\affiliation{\affA}\affiliation{\affB}
\author{R.~X.~Schüssler}\affiliation{\affA}

\newcommand{\MM}[1]{{\color{Orange}{#1}}}
\newcommand{\RG}[1]{{\color{Blue}{#1}}}
\newcommand{\RS}[1]{{\color{Red}{#1}}}
\newcommand{\CP}[1]{{\color{Violet}{#1}}}
\newcommand{\ND}[1]{{\color{Green}{#1}}}

\date{\today}% It is always \today, today,

\begin{abstract}

We present spectroscopic data for four metastable state clear-out transitions between 399\,nm and 412\,nm for all even long-lived isotopes of Yb$^+$ as well as their hyperfine structure in $^{171}$Yb$^+$. 
The strong \text{$^2 \rm{D}_{3/2} \rightarrow {}^1[1/2]_{1/2}$} transition at 412\,nm represents an attractive alternative for the standard 935\,nm repumper used in cooling and trapping experiments, while the transition to the \text{$^3[3/2]_{3/2}$} state at 411\,nm clears out the \text{$^2 $F$_{7/2}$} state, for which typically 638\,nm or 760\,nm are used.
These two alternative transitions simplify the experimental setup by removing the need for infrared optics to cool and manipulate Yb$^+$ and may be of particular interest when considering integrated photonics solutions. 
We also present data for the transitions \text{$^2 $D$_{3/2} \rightarrow {}^3[1/2]_{3/2}$} at 399\,nm,  and \text{$^2 $D$_{3/2} \rightarrow {}^1[5/2]_{5/2}$} at 410\,nm including decay branching ratios of the excited states.

\end{abstract}

\maketitle

\section{\label{sec:level1} Introduction}

The spectrum of Yb$^+$ has been studied extensively~\cite{Berends:1993,Pinnington:1994,Pinnington:1997,Li:1999}, motivated by its application in astronomy~\cite{Honda_2004,astronomy1,astronomy2,Sneden_2009}, as well as its broad use in quantum technologies~\cite{Olmschenk:2007,Monroe:2021}. Moreover, precise spectroscopic data is used to benchmark and improve atomic structure calculations~\cite{Fawcett:1991,Biemont:1998,Lea:2006,Safronova:2009,Dzuba:2011,Porsev:2012,Migdalek:2012,Roy:2017}. Rapid developments in trapping and laser cooling of single ions have greatly enhanced the precision of ion spectroscopy. In particular, using single laser cooled ions eliminates uncertainties due to Doppler broadening, background gas collisions and interactions between the ions. 
This has propelled the development of atomic clocks~\cite{Ludlow:2015,Huntemann:2016,Hausser:2024} and allowed tests of fundamental physics \cite{Peik:2004,dreissen_improved_2022,Hur:2022,door2024searchnewbosonsytterbium}. 
Yb$^+$ possesses a complex level structure owing to large configuration interactions and spin-orbit couplings. Theoretical predictions for the strongest optical transitions in Yb$^{+}$ have been published by Fawcett et al.~\cite{Fawcett:1991} and Bi\'emont et al.~\cite{Biemont:1998}. So far, most spectroscopic works on single Yb$^{+}$ ions have focused on the narrow, dipole-forbidden transitions~\cite{Taylor:1997,Blythe:2003,Schneider:2005,Furst:2020} and the dipole transitions used in laser cooling~\cite{Bell:1991,Taylor:1999,Olmschenk:2009,Meyer:2012,Feldker:2017}.
Many of the vast number of optical transitions have not been explored for their potential applications in laser cooling, trapping, and manipulation of the ions.

In this paper, we report spectroscopic data for four clear-out transitions that can be used to pump Yb$^+$ out of the metastable $^2$D$_{3/2}$ or $^2$F$_{7/2}$ states. 
We determine the transition frequencies for all stable even isotopes as well as the hyperfine splittings of $^{171}$Yb$^+$. 

\begin{figure*}
    \centering
    \includegraphics[trim={0 1cm 0 0}]{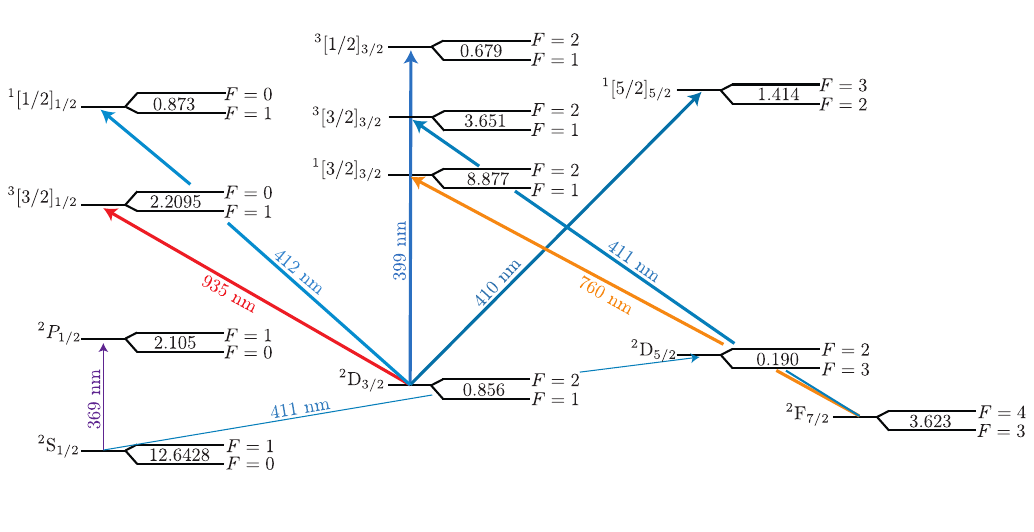}
    \caption{Level scheme for $^{171}$Yb$^{+}$: the transitions indicated in bold are measured in this work. The measured hyperfine splittings of these transitions are given in GHz with an uncertainty of 10\,MHz. The hyperfine splitting of the other states are also in GHz and are taken from:
    ${}^2\text{P}_{1/2}$ \cite{Olmschenk:2007}, ${}^2\text{D}_{5/2}$~\cite{Tan:2021}, ${}^2\text{S}_{1/2}$ \cite{Fisk:1997}.
    The hyperfine splittings of the ${}^2\text{D}_{3/2}$ and ${}^2\text{F}_{7/2}$ states are the average of all measured lines. } %${}^2\text{P}_{3/2}$ \cite{Feldker:2017}}
    \label{levelscheme}
\end{figure*}

A summary of our results is given in ~\autoref{levelscheme} and ~\autoref{tab:evenIsotopes} and \autoref{hyperfine_results}.
Furthermore, we investigate the usefulness of the studied transitions in laser cooling and state manipulation of trapped ions. 
We find that the \text{$^2 \rm{D}_{3/2} \rightarrow {}^1[1/2]_{1/2}$} transition at 412\,nm is very strong with a minimum linewidth of $\sim$~23(8)\,MHz at vanishing laser power and the transition is easily broadened to hundreds of MHz with modest laser power (see \autoref{sec:mccartney}). 
This transition offers a potential alternative to the 935\,nm transition that is more commonly used as a repumper~\cite{Bell:1991}. 

Additionally, we study the 411\,nm clear-out transition of the extremely long-lived $^2$F$_{7/2}$ state to the \text{$^3[3/2]_{3/2}$} state (see \autoref{sec:harrison}). 
Although we only have limited power available ($<3$\,mW) to drive this quadrupole transition, we find that the clear-out process is competitive with the more commonly used 760\,nm and 638\,nm solutions \cite{Taylor:1997}. 
Finally, we study the \text{$^2$D$_{3/2} \rightarrow {}^3[1/2]_{3/2}$} transition at 399\,nm (see \autoref{sec:lennon}) as well as the \text{$^2$D$_{3/2} \rightarrow {}^1[5/2]_{5/2}$}  transition at 410\,nm (see \autoref{sec:starr}) including the decay paths from the upper states. These results may be used to benchmark atomic structure calculations.

\section{Experimental Setup}
We trap a single Yb$^{+}$ ion in linear Paul trap. The trap is driven by a radio frequency field of 5.8\,MHz, resulting in secular trap frequencies of $\omega_{x,y,z}=2\pi\times(400, 400, 120)$\,kHz (see~\cite{Mazzanti:2023} for more details). Excess micromotion was compensated to amplitudes $\lesssim$~50\,nm. 

A level scheme of the relevant transitions for this work in $^{171}$Yb$^+$ is shown in \autoref{levelscheme}. The Yb$^+$ ions are produced by first heating up a sample of atomic Yb which is then ionized by a two-photon process in the trap center. 
For the first ionization step a 398.9\,nm laser is used and excites the ${}^1\text{S}_0\rightarrow{}^1\text{P}_1$ transition of the target isotope. 
A laser at 369\,nm is used for the second step of the ionisation process as well as for Doppler cooling on the ${}^2\text{S}_{1/2}\rightarrow{}^2\text{P}_{1/2}$ transition in Yb$^+$.
There is a weak decay path of 0.5\,\% from the ${}^2\text{P}_{1/2}$ state to the metastable ${}^2\text{D}_{3/2}$ state which has a lifetime of $\sim$52\,ms~\cite{Yu:2000,Schacht:2015}. Therefore, we use a 935\,nm laser to pump the population back to the cooling cycle via excitation to the ${}^3\text{[3/2]}_{1/2}$ state.
%52.7(2.4)\,ms \cite{Yu:2000}. 
Furthermore, the ion can also populate the long-lived ${}^2\text{F}_{7/2}$ state. We use a laser at 760\,nm  to excite the population to the ${}^1\text{[3/2]}_{3/2}$ state from which it can decay back to the cooling cycle.

\begin{table*}
\centering
\begin{tabular}{ccccccccccccccc}
\toprule
\multicolumn{2}{c}{Lower state}   &   \phantom{f}      &\multicolumn{7}{c}{\text{$^2 \text{D}_{3/2}$}  } & \phantom{f}& \multicolumn{3}{c}{\textrm{$^2 \text{F}_{7/2}$}} \\ 
%  & McC. & Len & Starr & &  &Harr\\
\multicolumn{2}{c}{Upper state}&           & \textrm{${}^1[1/2]_{1/2}$} &  & \textrm{${}^3[1/2]_{3/2}$} &   & \textrm{${}^1[5/2]_{5/2}$}& & \textrm{$^3[3/2]_{1/2}$}&  & \textrm{$^3[3/2]_{3/2}$}  & & \textrm{$^1[3/2]_{3/2}$} \\ \midrule
 &     Isotope        &  & 412\,nm &    & 399\,nm  &  & 410\,nm & & 935\,nm& \phantom{M}& 411 nm  && 760\,nm  \\ \midrule
&168& & 727\,533\,757&  & 751\,394\,594& & 730\,548\,033 &  & 320\,562\,140 & &729\,061\,398 & & 394\,432\,870 \\
&170           &      & 727\,537\,172 & & 751\,395\,899 & & 730\,550\,738 & & 320\,565\,870 & &729\,056\,942 & & 394\,429\,590\\
&172           &       & 727\,540\,360 & & 751\,397\,112 & &  730\,553\,260  & & 320\,569\,350   &  & 729\,052\,770 & & 394\,426\,550\\
&174            &      & 727\,542\,754 & & 751\,397\,858 & &  730\,555\,133 & &320\,571\,970 & &729\,049\,536& &394\,424\,150 \\
&176             &     & 727\,545\,028 & & 751\,398\,779 & & 730\,556\,908    &  &320\,574\,460 & &729\,046\,459 & & 394\,421\,890\\ \bottomrule
\end{tabular}%
\caption{Frequencies of the four investigated transitions for even isotopes.
For completeness we have included the measurement results for the clear-out transitions at 935\,nm and 760\,nm. 
All frequencies are given in MHz. 
The uncertainties are quoted to be 30\,MHz (for more details see text). 
For the four blue transitions the statistical uncertainties of the measurements are below 1 MHz. 
For the 935\,nm and 760\,nm transitions, the frequencies were determined were determined by a rough frequency scan documented only at the 10\,MHz level.}
\label{tab:evenIsotopes}
\end{table*}

\begin{table*}
\begin{tabular}{clclllllcccc}
\toprule
%McCartney &  &\phantom{d} &Lennon & & & Starr\\
\multicolumn{3}{c}{412\,nm } &\phantom{d} & \multicolumn{3}{c}{399\,nm }&\phantom{d} & \multicolumn{3}{c}{410\,nm }  \\
\multicolumn{3}{c}{$|{}^2\textrm{D}_{3/2},F\rangle\rightarrow|{}^1\textrm{[1/2]}_{1/2},F'\rangle$} &\phantom{d} & \multicolumn{3}{c}{$|{}^2\textrm{D}_{3/2},F\rangle\rightarrow|{}^3\textrm{[1/2]}_{3/2},F'\rangle$}&\phantom{d} & \multicolumn{3}{c}{$|{}^2\textrm{D}_{3/2},F\rangle\rightarrow|{}^1\textrm{[5/2]}_{5/2},F'\rangle$}  \\
%  412nm  &  412nm   &\phantom{d}  &  399nm  & 399 nm &\phantom{d}  & 410 nm    & 410 nm &    \\ 
\midrule
$F=2 \rightarrow F'=1 $ &   &   727\,537\,711 &  & $F=2\rightarrow F'=1$&   & 751\,395\,524&  &  $F=2 \rightarrow F'=2$ & &730\,550\,429  \\
$F=1 \rightarrow F'=1$  &  & 727\,538\,567 & &  $F=2\rightarrow F'=2$ &  &  751\,396\,206& &  $F=1 \rightarrow F'=2$  & &730\,551\,286   \\
$F=1 \rightarrow F'=0$  &  & 727\,539\,440&  & $F=1\rightarrow F'=1$&   & 751\,396\,380& &  $F=2 \rightarrow F'=3$  & &730\,551\,845  \\  
 $-$	 &    & $-$		&								 &$F=1\rightarrow F'=2$&  &  751\,397\,059&  & $-$	 & & $-$	 & \\
 \bottomrule \\
 \toprule
\multicolumn{3}{c}{935\,nm } &\phantom{d} & \multicolumn{3}{c}{760\,nm }&\phantom{d} & \multicolumn{3}{c}{411\,nm }  \\
\multicolumn{3}{c}{$|{}^2\textrm{D}_{3/2},F\rangle\rightarrow|{}^3\textrm{[3/2]}_{1/2},F'\rangle$} &\phantom{d} & \multicolumn{3}{c}{$|{}^2\textrm{F}_{7/2},F\rangle\rightarrow|{}^1\textrm{[3/2]}_{3/2},F'\rangle$}&\phantom{d} & \multicolumn{3}{c}{$|{}^2\textrm{F}_{7/2},F\rangle\rightarrow|{}^3\textrm{[3/2]}_{3/2},F'\rangle$}  \\
  \midrule
$F=2 \rightarrow F'=1$& &320\,566\,177& &$F=3 \rightarrow F'=1$ & & 394\,424\,969 & &$F=3\rightarrow F'=1$ & &  729\,055\,214\\
$F=1 \rightarrow F'=1$ &  &320\,567\,035 & & $F=4 \rightarrow F'=2$ & & 394\,430\,225 & &$F=4\rightarrow F'=2$ & & 729\,055\,244 \\
$F=1 \rightarrow F'=0$ & &320\,569\,244& &$F=3 \rightarrow F'=2$ & &  394\,433\,848& & $F=3\rightarrow F'=2$& & 729\,058\,866\\
\bottomrule
\end{tabular}%
\caption{Frequencies of the hyperfine transitions in ${}^{171}$Yb${}^{+}$ for the investigated transitions. For completeness we have included the measurement results for the clear-out transitions at 935\,nm and 760\,nm. All frequencies given in MHz. Uncertainties are quoted to be 30\,MHz, while statistical uncertainties from a Lorentzian fit to the data are all below 1\,MHz.}%935 F=2 $\rightarrow$ F=1: Folder06, the other two from their respective folder 04.}
\label{hyperfine_results}
\end{table*}

In order to cool and image ${}^{171}$Yb$^{+}$, the hyperfine splittings of the transitions need to be bridged.
We use a 14.7\,GHz electro-optic modulator (EOM) for the 369\,nm Doppler transition, a 3.07\,GHz fiber EOM for the 935\,nm repumper, and a 5.26\,GHz fiber EOM for the 760\,nm clear-out transition. 
Microwaves at 12.6\,GHz to bridge the ground state hyperfine splitting can also be supplied using a microwave horn~\cite{Olmschenk:2007}.

We image the ion's fluorescence at 369\,nm wavelength on an EMCCD camera. 
The 935\,nm and 760\,nm lasers are aligned along the axial axis of the linear trap. A magnetic field of $\sim$~0.3\,mT, oriented perpendicular to the trap axis, is used to avoid coherent population trapping during Doppler cooling. The 369\,nm laser and the blue spectroscopy lasers enter at an angle of 45$^{\circ}$ to the axial direction and 90$^{\circ}$ to the magnetic field. We avoid the use of circular laser polarization to minimize optical pumping effects that may cause line shifts. 

Two lasers are used to measure the four transitions. 
The frequency of the three studied transitions spanning the 410-412\,nm range are easily reached by tuning the wavelength of the 411\,nm laser that is normally used for driving the ${}^2\text{S}_{1/2}\rightarrow{}^2\text{D}_{5/2}$ clock transition \cite{Hirzler2020esf}. 
The 399\,nm transition from ${}^2\text{D}_{3/2}\rightarrow{}^3\text{[1/2]}_{3/2}$ is reached with a home-built laser based on \cite{Ewald:2015}.
It is noteworthy that this transition is only $\sim$~130\,GHz away from the first step ionization laser that drives the $^1$S$_0\rightarrow ^1$P$_1$ transition in neutral Yb.

\section{Results}

All lasers are locked to a wavemeter (High Finesse WS8-10) using a computer controlled feedback loop.
The wavemeter is calibrated with a 638\,nm laser from a neighbouring laboratory locked to the ${}^1\text{S}_{0}\rightarrow{}^3\text{P}_{1}$ clock transition in ${}^{88}$Sr at 434.829\,121\,300(20)\,THz \cite{PhysRevResearch.4.023245,courtillot_accurate_2005}.
This laser has a linewdith of 1.2\,kHz and drifts about 10\,kHz per day.
The wavemeter calibration is quoted to be valid for 1\,h with an error of $10$\,MHz for wavelengths within a $\pm$200\,nm range around the calibration wavelength.
Outside of this range, where the presented results fall, the measured values are shifted and the uncertainty is stated to be 30\,MHz \cite{highfinesse}. 
To estimate this systematic shift, we measured the ${}^2\text{S}_{1/2}\rightarrow{}^2\text{D}_{5/2}$ clock transition for $^{174}$Yb$^+$ several times over the span of two weeks to be 729\,475\,284(2)$_{\textrm{stat}}$\,MHz which differs by only $\sim$0.2\,MHz from the best known value~\cite{door2024searchnewbosonsytterbium}.
Therefore, we expect our results to be more precise than the quoted uncertainty of the 30\,MHz.
We quote an uncertainty of 10\,MHz for the hyperfine splittings, but a more conservative 30\,MHz uncertainty for the frequencies of the optical transitions.

\begin{figure}
    \centering
    \includegraphics[trim={1.3cm 0 0 0},width=0.95\linewidth]{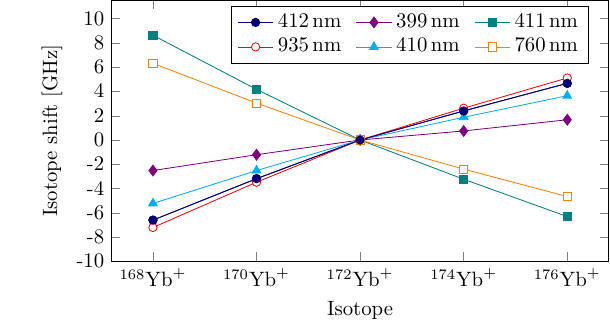}
    \caption{The isotope shifts for the six measured transitions with the result for $^{172}$Yb$^+$ taken as the reference. 
    The uncertainty of 30\,MHz is smaller than the marker size.
    }
    \label{isotopeshift}
\end{figure}

The frequencies of the transitions were determined by scanning the lasers on a MHz level while simultaneously collecting the fluorescence of a single trapped ion on the camera for a set detection time.
For each single fluorescence measurement the frequency of the laser was read out using the wavemeter, i.e. our data consists of unique laser frequencies paired with a number of detected photons. 

For all data with the exception of the 412\,nm and 935\,nm transitions, a pulsed scheme was used to measure the frequencies. Here, the fluorescence data was analysed by projecting it on one of two outcomes.
A threshold was set on the measured photon counts to distinguish the cases where the ion was in the bright ($^2$S$_{1/2}$, $^2$D$_{3/2}$) state and the dark ($^2$F$_{7/2}$)  state. Owing to the long lifetime of the dark state, the detection error can be made negligible by increasing the detection time~\cite{Edmunds:2021}.
From the Lorentzian fit to the  data we get typical statistical uncertainties of $\sim$200\,kHz for the frequencies with outliers of 700\,kHz for the low-abundant ${^{168}}$Yb$^{+}$. These are negligible compared to the wavemeter uncertainty and we quote the latter as the dominant error. The hyperfine splittings in ${^{171}}$Yb$^{+}$ were extracted from a combined fit to all transitions.

\subsection{${}^2\text{D}_{3/2}\rightarrow{}^1\text{[1/2]}_{1/2}$ at 412\,nm}
\label{sec:mccartney}
\begin{figure}[h]
    \centering
    \includegraphics[width=\linewidth]{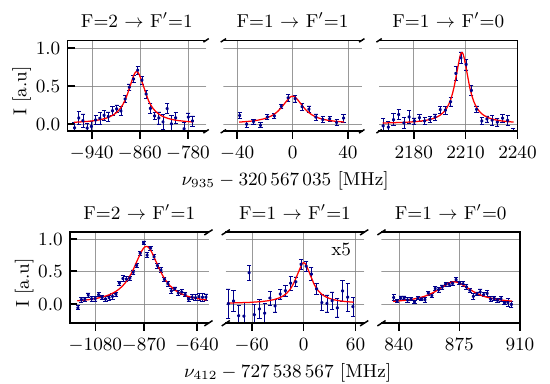}
    \caption{Ion fluorescence (I) as a function of frequency for the hyperfine structure of the levels involved in the ${}^2\text{D}_{3/2}\rightarrow{}^3\text{[3/2]}_{1/2}$ transition at 935\,nm (top) and the ${}^2\text{D}_{3/2}\rightarrow{}^1\text{[1/2]}_{1/2}$ transition at 412\,nm (bottom). In each set, the peaks are normalised to the largest peak and the data points represent an average of 4\,MHz bins. The amplitude of the transition from F = 1 $\rightarrow$ F$^{\prime}$ = 1 in the bottom graph is scaled by a factor of 5 to enhance its visibility relative to the other peaks. The red line indicates a Lorentzian fit, for more details see~\autoref{sec:mccartney}.}
    \label{412_935}
\end{figure}

We used the 369\,nm laser and scanned the spectroscopy laser around the predicted value of 412\,nm ~\cite{Fawcett:1991,Biemont:1998}. On resonance, the 412\,nm acts as a repumper for the $^2$D$_{3/2}$ state, resulting in a peak in ion fluorescence. We recorded the ion fluorescence for 100\,ms for each measurement. 
Such a dataset is shown in \autoref{412_935} for the hyperfine transitions, with the data averaged in 4\,MHz bins for clarity. 
For ${}^{171}$Yb${}^{+}$, we observe three peaks in the fluorescence spectrum, corresponding to the three dipole-allowed hyperfine transitions. To enhance the signal intensity for the $|^2$D$_{3/2} , F=2\rangle\rightarrow{}|^1[1/2]_{1/2},F=1\rangle$ transition, we tuned the 935\,nm laser on the $|^2$D$_{3/2} , F=1\rangle\rightarrow{}|^3[3/2]_{1/2},F=1\rangle$ transition to assist in pumping the population to the $|^2$D$_{3/2} , F=2\rangle$ state. 
The two lowest frequencies have a difference of 858(10)\,MHz, in agreement with the known hyperfine splitting of the $^2$D$_{3/2} $ state \cite{Olmschenk:2007} and we identify them as the $|^2$D$_{3/2} , F=2\rangle\rightarrow{}|^1[1/2]_{1/2},F=1\rangle$ and  $|^2$D$_{3/2} , F=1\rangle\rightarrow{}|^1[1/2]_{1/2}, F=1\rangle$ transitions.
The third peak is 873(10)\,MHz higher and we attribute it to the $|^2$D$_{3/2},F=1\rangle\rightarrow{}|^1[1/2]_{1/2},F=0\rangle$ transition. 
It is noted that the hyperfine structure of the $^1[1/2]_{1/2}$ state is inverted.

For completeness and comparison, we performed the same measurement procedure on the $^3$D$_{3/2}\rightarrow{}^3[3/2]_{1/2} $ transition using the 935\,nm laser instead of the 412\,nm laser. We used  For ${}^{171}$Yb${}^{+}$, we again find three fluorescence peaks as shown in~\autoref{412_935}. 
The difference between the three peaks is 858(10)\,MHz and 2209(10)\,MHz, which is in agreement with literature values \cite{Olmschenk:2007} for the hyperfine splittings of the two states.

\begin{figure}
    \centering
    \includegraphics[width=\linewidth]{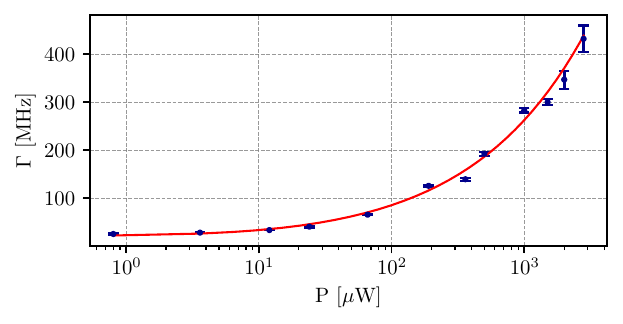}
    \caption{Linewidth as a function of laser power for the ${}^2\text{D}_{3/2}\rightarrow{}^1\text{[1/2]}_{1/2}$ transition at 412\,nm. The data is fitted with a Lorentzian function and gives an extrapolated linewidth, for vanishing laser power, of $\Gamma = 23(8)$\,MHz. For these measurements, the 369\,nm Doppler laser's power was set to  saturate the cooling transition.}
    \label{linewidth}
\end{figure}

For the 412\,nm transition, laser powers as low as 1\,µW with an estimated beam waist of \SI{390(40)}{\micro\meter} were enough to achieve repumping rates that are competitive with the 935\,nm beam at 200\,µW with a beam waist of \SI{360(30)}{\micro\meter} for the even isotopes.
We measured the transition linewidth of the 412 transition for different laser powers in  $^{174}$Yb$^{+}$, as depicted in \autoref{linewidth}.  Extrapolating to zero laser power, we estimate an upper limit of the natural linewidth of 23(8)\,MHz.
Remarkably, at only a few mW of power, the linewidth is broadened to several hundreds of MHz. 
In our setup, $\sim$~11\,mW should be enough to have a linewidth similar to the hyperfine splitting of the ${}^2\text{D}_{3/2}$ state in ${}^{171}$Yb${}^{+}$, eliminating the need for an EOM.

\subsection{${}^2\text{D}_{3/2}\rightarrow{}^3\text{[1/2]}_{3/2}$ at 399\,nm}
\label{sec:lennon}

The ion was prepared in the $^2$D$_{3/2}$ state using a 100\,µs pulse of 369\,nm light. 
A 2\,ms spectroscopy pulse at 399\,nm then excited the ion to the $^3\text{[1/2]}_{3/2}$ state, followed by a 20\,ms fluorescence measurement using the 369\,nm and 935\,nm lasers.
The $^3\text{[1/2]}_{3/2}$ state has a dipole-allowed decay channel to the $^2$D$_{5/2}$ state from which it can subsequently decay to the long-lived $^2$F$_{7/2}$ state, that remains dark during fluorescence detection. We identified the resonance of the 399\,nm transition with a reduction in ion fluorescence. After fluorescence detection, the 760\,nm laser cleared out the $^2$F$_{7/2}$ state before the start of the next measurement.

In order to determine the branching ratio, we added a wait time of 50\,ms after the 399 nm pulse, to let the $^2$D$_{5/2}$ state decay. We assume there is no significant direct decay on the $^3\text{[1/2]}_{3/2}\rightarrow ^2$F$_{7/2}$ quadrupole transition. Considering the branching ratio of 17\,\% of $^2$D$_{5/2}$ to $^2$S$_{1/2}$ state \cite{Taylor:1997}, we find the branching ratio between the decay paths  $^3\text{[1/2]}_{3/2}\rightarrow ^2$S$_{1/2}$ and $^3\text{[1/2]}_{3/2}\rightarrow ^2$D$_{5/2}$ to be 8.4(1.4).

\begin{figure}[t]
    \centering
    \includegraphics[width=\linewidth]{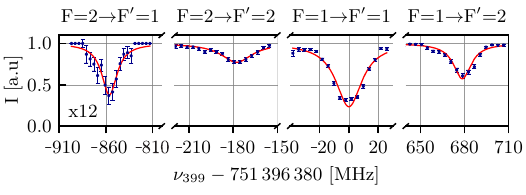}
    \caption{Ion fluorescence (I) as a function of frequency for the hyperfine structure of the levels involved in the ${}^2\text{D}_{3/2}\rightarrow{}^3\text{[1/2]}_{3/2}$ transition at 399\,nm. The data points represent an average of 4 MHz bins. The amplitude of the transition from F = 2 $\rightarrow$ F$^{\prime}$ = 1 is scaled by a factor of 12 to enhance its visibility relative to the other dips. The red line indicates a Lorentzian fit, for more details see \autoref{sec:lennon}. }
    \label{lennon}
\end{figure}

For measuring the hyperfine structure in ${}^{171}$Yb${}^{+}$, we increased the spectroscopy pulse of 399\,nm to 50 ms and kept the 369\,nm and 935 \,nm lasers on. The latter was tuned to clear out the hyperfine level that is not targeted by the 399\,nm laser. 
We observe four distinct dips, as shown in \autoref{lennon}. 
We find two sets of frequencies that are 854(10)\,MHz apart, corresponding to $^2$D$_{3/2}$ hyperfine splitting. 
The frequency difference of the other two sets of 679(10)\,MHz gives the splitting of the $^3[1/2]_{3/2}$ state.

The ordering of the hyperfine levels of $^3[1/2]_{3/2}$ was found by determining the hyperfine level of the $^2$F$_{7/2}$ state that the ion decayed to. 
From the $|^3[1/2]_{3/2},F=1\rangle$ state, it can only decay to $|^2$D$_{5/2},F=2\rangle$ due to selection rules.
When it does not decay to the $^2$S$_{1/2}$ ground state, the $|^2$D$_{5/2},F=2\rangle$ state can only decay to the $|^2$F$_{7/2},F=3\rangle$ state while the $|^2$D$_{5/2},F=3\rangle$ state predominantly decays to $|^2$F$_{7/2},F=4\rangle$. 
We determined the final hyperfine level of the $^2$F$_{7/2}$ state that the ion decayed to using the 760\,nm transition, for which all the hyperfine splitting and levels are known. 
The assignment of the transitions is shown in \autoref{lennon}. We note that the hyperfine structure of the $^3[1/2]_{3/2}$ state is not inverted.

\subsection{ ${}^2\text{D}_{3/2}\rightarrow{}^1\text{[5/2]}_{5/2}$ at 410\,nm}
\label{sec:starr}

We used a pulsed scheme similar to \autoref{sec:lennon}, where we pumped the ion to the $^2$D$_{3/2}$ state for 100\,µs and then performed a 5\,ms spectroscopy pulse of 410\,nm light to excite the ion to the $^1\text{[5/2]}_{5/2}$ state.
This state can decay to the $^2$D$_{5/2}$ state and subsequently to the $^2$F$_{7/2}$ state. As before, we measured a reduction in fluorescence when the 410\,nm laser was on resonance.

\begin{figure}
    \centering
    \includegraphics[width=\linewidth]{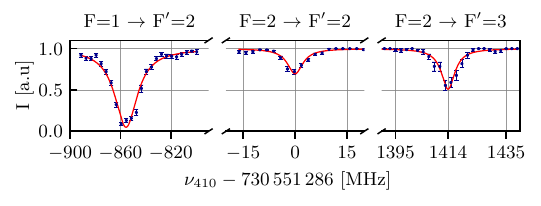}
    \caption{Ion fluorescence (I) as a function of frequency for the levels involved in the ${}^2\text{D}_{3/2}\rightarrow{}^1\text{[5/2]}_{5/2}$ transition at 410\,nm. The data points represent an average of 4 MHz bins. The red line indicates a Lorentzian fit, for more details see \autoref{sec:starr}.}
    \label{starr}
\end{figure}

For ${}^{171}$Yb${}^{+}$, we find three dips of fluorescence.
Since the smallest frequencies are separated by 856(10)\,MHz, we can identify them as the  $|^2D_{3/2},F=1\rangle\rightarrow{}|^1[5/2]_{5/2},F=2\rangle$ and $|^2D_{3/2},F=2\rangle\rightarrow{}|^1[5/2]_{5/2},F=2\rangle$ transitions.
%One of the dips is separate by 860() MHz which we attribute to the hyperfine splitting of the $^2D_{3/2}$ state. 
The third frequency corresponds to $|^2D_{3/2},F=2\rangle\rightarrow{}|^1[5/2]_{5/2},F=3\rangle$, giving a hyperfine splitting of 1414(10)\,MHz for the $^1[5/2]_{5/2}$ state. The data is shown in \autoref{starr}.
The hyperfine structure of the $^1[5/2]_{5/2}$ state is not inverted.
 
To check that there is no direct decay on the $^1[5/2]_{5/2} \rightarrow{}^2$F$_{7/2}$ quadrupole transition, we perform a branching ratio measurement with $^{174}$Yb$^+$. 
For this we used a 2\,ms pulse to pump the ion in the $^2$D$_{3/2}$ state, followed by a 10\,ms pulse of 410\,nm light.
We detected the fluorescence after waiting 50\,ms, allowing all population in the $^2$D$_{5/2}$ state to decay.
We observed the ion in the ground state in 18(1)\,\% of the cases, for all pumping times $>$300\,µs and independent of the pulse length of 410\,nm light.
This result is in agreement with the known branching ratio of the $^2$D$_{5/2}$ state  \cite{Taylor:1997}. We therefore do not find measurable decay on the $^1[5/2]_{5/2}\rightarrow ^2$F$_{7/2}$ quadrupole transition.

\subsection{ ${}^2\text{F}_{7/2}\rightarrow{}^3\text{[3/2]}_{3/2}$ at 411\,nm}
\label{sec:harrison}

\begin{figure}
    \centering
    \includegraphics[width=\linewidth]{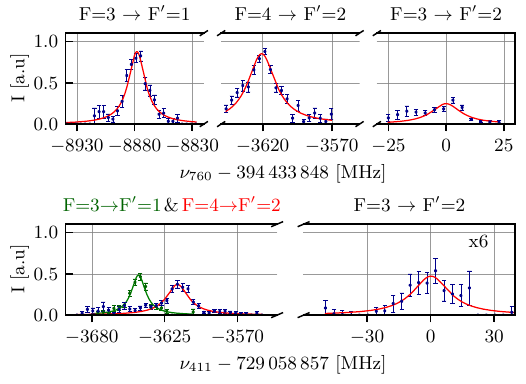}
    \caption{Ion fluorescence (I) as a function of frequency for the hyperfine structure of the levels involved in the  ${}^2\text{F}_{7/2}\rightarrow{}^1\text{[3/2]}_{3/2}$ transition at 760\,nm (top) and ${}^2\text{F}_{7/2}\rightarrow{}^3\text{[3/2]}_{3/2}$ transition at 411\,nm (bottom). The data points represent an average of 4 MHz bins. The amplitude of the transition from $\text{F}=3 \rightarrow \text{F}^{\prime} = 2$ in the bottom graphs is scaled by a factor of 6 to enhance its visibility relative to the other peaks. The red line indicates a Lorentzian fit, for more details see~\autoref{sec:harrison}.}
    \label{rerepumper}
\end{figure}

To measure the transition frequency, we prepared the ion in the $^2$F$_{7/2}$ state using a 30\,ms pulse of 369\,nm and 399\,nm light, see \autoref{sec:lennon}.
A pulse of 411\,nm light then excited the ion to the $^3[3/2]_{3/2}$ state, from where it could decay back to the bright states, resulting in a fluorescence peak when the 411 nm laser is resonant.

We measured the hyperfine structure of ${}^{171}$Yb${}^{+}$ using the 399\,nm laser, discussed in section~\autoref{sec:lennon}, to pump the ion to the $|^2$F$_{7/2},F=3\rangle$ state. The frequency difference of 3651(10)\,MHz between the two observed transitions corresponds to the hyperfine splitting of the $^3\text{[3/2]}_{3/2}$ state.
Additionally, by exciting the ion to the $|^2F_{7/2},F=4\rangle$ state and measuring the third transition, we were able to assign the hyperfine levels. We measure a difference of 3622(10)\,MHz to the higher of the other two frequencies, in agreement with the hyperfine splitting of the $^2F_{7/2}$ state~\cite{Taylor:1999}. We note that the hyperfine structure of the $^3\text{[3/2]}_{3/2}$ state is not inverted. 

For completeness and in order to enable comparison, we measured the hyperfine splittings of the 760\,nm transition. Here, the roles of the 760\,nm and 411\,nm lasers were reversed compared to the previous measurement. 
The three measured peaks, shown in \autoref{rerepumper}, give the known hyperfine splitting of the $^2$F$_{7/2}$ state.
The splitting of the upper state $^3[3/2]_{3/2}$ was measured to be 8877(10)\,MHz, in agreement with previous values \cite{Roman:2021}. 

With the 411\,nm laser we were limited to around $\sim$~2.3\,mW laser power with a waist of \SI{390(40)}{\micro\meter}, leading to a slower clear-out process than for the 760\,nm laser, for which we had 55\,mW available with a waist of \SI{360(30)}{\micro\meter}. 
This is reflected in the maximum repump efficiency of 50\,\% during a 100\,ms pulse as shown in  the hyperfine measurement, see \autoref{rerepumper}. This could easily be improved by using more laser power.

\section{Conclusion}
We have presented spectroscopic data for four clear-out transitions in Yb$^+$. The strong \text{$^2 $D$_{3/2} \rightarrow {}^1[1/2]_{1/2}$} transition at 412\,nm forms an attractive alternative for the standard 935\,nm repumper, while the transition at 410\,nm clears out the \text{$^2 $F$_{7/2}$} state via the \text{$^3[3/2]_{3/2}$} state.
We note that it is quite remarkable that all transitions that are needed for photoionizing Yb, laser cooling and manipulation of the ions fall within the narrow wavelength window of 369-412\,nm.
This property is not shared by the simpler heavy earth-alkaline ions Ca$^+$, Sr$^+$ and Ba$^+$ that have metastable states and are often used in cooling and trapping experiments. 
This feature may be of practical use when considering integration of optical elements in micro fabricated ions traps which would be simpler to produce for a narrow range of wavelengths \cite{kwon_multi-site_2024}.

In the future, further narrowing of the wavelength window would be achieved by studying the $^2$D$_{3/2}\rightarrow{}^3[1/2]_{1/2}$ clear-out transition at 378\,nm, as well as the $^2F_{7/2}\rightarrow{}^3[3/2]_{5/2}$ and $^2F_{7/2}\rightarrow{}^3[1/2]_{3/2}$ clear-out transitions at 372 and 376\,nm, respectively.

\section*{Acknowledgements}
We thank the group of Florian Schreck for making available the Sr clock laser for wavelength calibration. This work was supported by the Dutch Research Council (Grant Nos. 680.91.120, VI.C.202.051 and 680.92.18.05, R.G., M.M. and R.X.S.). 

\setcounter{equation}{0}
\setcounter{figure}{0}
\setcounter{table}{0}

\bibliography{paper}

\end{document}